\documentclass{iopart}
\usepackage{iopams}
\usepackage{amssymb,amstext,amsfonts}
\usepackage{graphicx}
\usepackage[sort&compress,comma,square]{natbib}

\usepackage{color}

\newcommand{\rem}[1]{}

\begin{document}

\title[From Equilibrium to Steady State]
{From Equilibrium to Steady State: The Transient Dynamics of Colloidal Liquids under Shear} 
\author{J Zausch$^{1}$, J Horbach$^{1,2}$, M Laurati$^{3}$, S U Egelhaaf$^{3}$, 
J M Brader$^{4}$, Th Voigtmann$^{2,4}$, and M Fuchs$^{4}$}
\address{$^{1}$Institut f\"ur Physik, Johannes-Gutenberg-Universit\"at Mainz, 
               Staudinger Weg 7, 55099 Mainz, Germany\\
         $^{2}$Institut f\"ur Materialphysik im Weltraum, Deutsches Zentrum f\"ur
               Luft- und Raumfahrt (DLR), 51170 K\"oln, Germany\\
         $^{3}$Condensed Matter Physics Laboratory,
              Heinrich-Heine-Universit\"at, Universit\"atstra\ss e 1, 40225 
               D\"usseldorf, Germany\\  
         $^{4}$Fachbereich Physik, Universit\"at Konstanz, 78457 Konstanz, Germany}
\date{\today}

\def\eqref#1{(\ref{#1})}
\let\bs\boldsymbol

\newcommand{\tlname}[1]{\ensuremath{\text{\textit{#1}}}}
\def\Pe{\tlname{Pe}$_0$}
\def\We{\tlname{\rm Pe}}
\renewcommand{\vec}[1]{\bf{#1}}
\begin{abstract}
We investigate stresses and particle motion during the start up of
flow in a colloidal dispersion close to arrest into a glassy state.
A combination of molecular dynamics simulation, mode coupling theory
and confocal microscopy experiment is used to investigate the origins
of the widely observed stress overshoot and (previously not reported)
super-diffusive motion in the transient dynamics. A  link between
the macro-rheological stress versus strain curves and the microscopic
particle motion is established. Negative correlations in the transient
auto-correlation function of the potential stresses are found responsible
for both phenomena, and arise even for homogeneous flows and almost
Gaussian particle displacements.
\end{abstract}
%


%
\section{Introduction}
Relaxation processes in highly viscous liquids are strongly affected by
the application of an external shear field. Even at small shear rates, the
shear viscosity can decrease by orders of magnitude while the structure
exhibits only small changes compared to equilibrium. 
This phenomenon of shear thinning has been observed in experiments
of various systems, such as colloids and polymers, as well as in
computer simulations of model systems \cite{larson}.
The transport coefficients measured in quiescent equilibrium or
linear response markedly differ from those in steady-shear states.
This raises the question how the system's microscopic dynamics evolves
in response to a sudden change in the externally applied shear.
This concerns the transient dynamics, as, for example, displayed in
time-dependent transport coefficients or waiting-time dependent two-time
correlation functions.  Here, we are concerned with a suddenly commencing
shear flow, imposed upon a quiescent, equilibrated system. While the
measurement of macroscopic quantities like the shear stress $\sigma$ is
standard rheology procedure, the detailed study of microscopic dynamics
has been a recent contribution to the field from both computer simulation
\cite{Yamamoto1998,Strating1999,Berthier2002,Varnik2003,Varnik2004,Varnik2006,Varnik2006b,mcphie06}
and direct-imaging techniques for colloidal suspensions, namely confocal
microscopy \cite{Besseling2007}.

It is well known that after starting a steady shear flow, stresses develop
in the liquid. At short times after switching on (corresponding to small
overall strains $\gamma=\dot\gamma t$), they increase almost linearly
and at long times saturate at a constant level, the stress corresponding
to the sheared steady state. In many cases, the increase in stress first
continues beyond this steady-state value before relaxing back at larger
times. A nonmonotonic $\sigma(t)$-versus-$t$ curve results, with a maximum
at intermediate strains, the `stress overshoot'. Such overshoots have
been seen in polymeric liquids \cite{Osaki2000,Islam2001,overshoot},
metallic glasses \cite{aken00,johnson02}, and in simulations
\cite{heyes86,utz99,Rottler2003,Varnik2004,tanguy06}. Despite its
abundance among different systems, no generally accepted physical
picture has emerged. Linking this macroscopically observed feature
to the underlying microscopic dynamics can be achieved by studying
the transient (microscopic) correlation functions. We pursue this by
combining confocal microscopy, molecular-dynamics simulation, and a
mode-coupling-theory approach. Surprisingly, we find that a previously
not reported, super-diffusive particle motion is characteristic for the
dynamics at strain values where the stress overshoot appears. We connect
both phenomena to negative portions in the transient auto-correlation
function of the microscopic stresses.

The mode-coupling theory (MCT) for colloidal rheology
\cite{fuchs02,fuchs05,faraday,brader07,brader08pre} is a recent
microscopic framework from which constitutive equations can be derived
that capture the interplay of slow dynamics and shear thinning. The
theory can thus address the above question on the relation between
microscopic dynamics and macroscopic rheology.  Its underlying
physical picture is a combination of the cage effect, leading to
slow dynamics, and of the advection of long-wavelength fluctuations
to short wavelength, leading to a shear-induced breaking of cages
\cite{miyazaki04,miyazaki07,wyss07,szamel04}. The MCT equations can be
formulated from an integration-through-transient formalism developed by
Fuchs and Cates \cite{fuchs02,fuchs05}. In this formalism, transient
correlation functions of a special kind (equilibrium averages over
fluctuations that evolve according to the nonequilibrium dynamics) play
a central r\^ole. The direct measurement of such correlations, which is
possible under shear startup, provides a direct link between theory,
simulation and experiment. This allows for a detailed test of some of
the assumptions implicit in the theory.

The paper is structured as follows: we will in
section~\ref{sec:techniques} discuss the different techniques used
in this study, viz.\ the simulation model, experimental setup, and
theoretical framework. Section~\ref{sec:results} presents and compares
the main results for both stress buildup and waiting-time dependent
mean-squared displacements after switching on shear flow. Finally,
section~\ref{sec:conclusions} concludes.

\section{Methods and Techniques}\label{sec:techniques}
\subsection{Simulation model and details of the simulation}
Molecular dynamics computer simulations have been done for a simple
model of a binary AB mixture of charged colloids. The interactions
between the particles are modeled by a Yukawa potential,
\begin{equation}
\label{eq_pot}
u_{\alpha\beta}=
\epsilon_{\alpha \beta} d_{\alpha \beta}
\frac{\exp(- \kappa_{ \alpha \beta}(r- d_{\alpha\beta}))}{r}
\quad \alpha,\beta=\rm{A,B},
\end{equation}
which is truncated at a cut-off distance $r_{\rm
c}^{\alpha\beta}$, defined by $u_{\alpha\beta}(r_{\rm
c}^{\alpha\beta})=10^{-7}\,\epsilon_{\rm AA}$.  The ``particle diameters''
are set to $d\equiv d_{\rm AA}=1.0$, $d_{\rm BB}=1.2\,d$ and $d_{\rm
AB}=1.1\,d$, the energy parameters to $\epsilon\equiv\epsilon_{\rm
AA}=1.0$, $\epsilon_{\rm BB}=2.0\,\epsilon$, $\epsilon_{\rm
AB}=1.4\,\epsilon$, and the screening parameters to $\kappa_{\rm
AA}=\kappa_{\rm BB}=\kappa_{\rm AB}=6/d$. The choice of these parameters
ensures that, at the density $\varrho=0.675\,m_{\rm A}/d_{\rm
AA}^3$ considered in this work, no problems with crystallization
or phase-separation occur, at least in the temperature range under
consideration.  The masses of the particles are set to unity, i.e.~$m=m_{\rm
A}=m_{\rm B}=1.0$.

The simulations were done for a 50:50 mixture of $N=2N_{\rm A}=2N_{\rm
B}=1600$ particles, placed in a cubic simulation box of linear size
$L=13.3\,d$. For the sheared system, we chose the $x$ direction as the
direction of shear and the $y$ and $z$ direction as the gradient and
vorticity direction, respectively.  Shear was imposed onto the system
via modified periodic boundary conditions, the so-called Lees-Edwards
boundary conditions \cite{lees72,evansbook}.  Here, a particle that moves
out of the simulation box in $y$ direction is subject to a displacement
in $x$ direction due to constant velocities $u_{{\rm s},x}$ and $-u_{{\rm s},x}$ of the
image cells above and below the simulation cell, respectively.  For the
glassforming Yukawa system considered in this work, the application
of Lees-Edwards boundary conditions leads to a linear shear profile
in the steady state regime, $v_{{\rm s},x}(y)=\dot{\gamma} (y-L/2)$ with the shear
rate $\dot{\gamma}=u_{{\rm s},x}/L$.  In recent simulation studies, Lees-Edwards
boundary conditions have been used in conjunction with the so-called
Sllod equations of motion \cite{evansbook} to enforce the formation of
a linear shear profile. For our purpose, the use of the Sllod equations
is not appropriate since we are mainly interested in the study of the
transient dynamics, i.e.~in the time regime before the steady state
regime is reached. In this case, one may expect that the emergence of
non-linear shear profiles strongly affects the response of the system
to the external shear field. Thus, the Sllod equations would modify the
transient dynamics in an artificial manner. However, as we shall see
below, an almost linear profile is built up long before the steady state
regime is approached. So our simulations indicate {\it a posteriori}
that it does not matter whether or not the Sllod equations are used. We
also note that non-linear shear profiles may occur as a consequence
of a hydrodynamic instability, leading to the formation of shear bands.
However, in the glassforming Yukawa system of this work, such phenomena
are not observed, at least for the considered range of shear rates.

Both for the equilibrium simulations and the simulations under shear, the
system was coupled to a dissipative particle dynamics (DPD) thermostat
(see Ref.~\cite{soddemann03} and references therein). The DPD equations
of motion are given by
\begin{equation}
\dot{{\bf r}}_i=
\frac{{\bf p}_i}{m_i},\quad
\dot{{\bf p}_i}=\sum_{j(\neq i)}
   \left[{\bf F}_{ij}+{\bf F}^{\rm D}_{ij}+{\bf F}^{\rm R}_{ij}\right]
\label{eq_dpd}
\end{equation}
with ${\bf r}_i$ and ${\bf p}_i$ the position and momenta
of a particle $i$ ($i=1,..., N$).  In Eq.~(\ref{eq_dpd})
${\bf F}_{ij}=-{\nabla}u_{ij}$ denotes the conservative force between a
particle $i$ and a particle $j$ due to the interaction potential defined
by Eq.~(\ref{eq_pot}). To provide the thermostatting of the system a
dissipative force ${\bf F}^{\rm D}_{ij}$ and a random force ${\bf F}^{\rm
R}_{ij}$ are added in Eq.~(\ref{eq_dpd}).

The dissipative force is defined by \cite{espanol95}
\begin{equation}
\label{eq_dissf}
  {\bf F}^{\rm D}_{ij}=-\zeta\,w^2(r_{ij})\,
\left(\hat{{\bf r}}_{ij}\cdot {\bf v}_{ij}\right)\,\hat{{\bf r}}_{ij}  
\end{equation}
with $\zeta$ being a friction coefficient,
${\bf v}_{ij}={\bf v}_i-{\bf v}_j$ the relative velocity between particle
$i$ and $j$, $\hat{{\bf r}}_{ij}$ the unit vector of the vector ${\bf r}_{ij}=
{\bf r}_i - {\bf r}_j$, and $r_{ij}=| {\bf r}_i - {\bf r}_j |$ the
distance between particle $i$ and $j$. The function $w(r_{ij})$ is defined
by $w(r_{ij})=\sqrt{1-r_{ij}/r_c}$ for $r_{ij}<r_c^{\rm DPD}=1.25\,d$
and $w(r_{ij})=0$ otherwise. Thus, the force ${\bf F}^{\rm D}$ describes
a frictional force due to the interaction between neighboring particle
pairs. The use of the relative velocities between neighboring particles
in (\ref{eq_dissf}) is crucial to obtain the correct behaviour on hydrodynamic
scales. It ensures that the DPD thermostat is Galilean invariant and
that momentum is locally conserved. For the friction coefficient $\zeta$
we chose the value $\zeta=12$. With this value, the microscopic properties
are close to those of a purely Newtonian dynamics with $\zeta=0$.  The
microscopic dynamics can be significantly changed by using high values of
$\zeta$, approaching the limit of an overdamped stochastic dynamics
where inertia effects can be effectively neglected. In a forthcoming
publication \cite{zauschpre}, we present simulations with $\zeta=1200$
and compare them to those with $\zeta=12$ that are shown in the present
paper. These simulations indicate that qualitatively all the essential
features of the transient dynamics do not depend on the choice of $\zeta$.

The random force in (\ref{eq_dpd}) is given by ${\bf F}^{\rm
R}_{ij}=\sqrt{2k_B T\zeta}\,w(r_{ij})\,\theta_{ij}\, \hat{{\bf r}}_{ij}$
where $\theta_{ij}=\theta_{ji}$ are uniform random numbers with zero
mean and unit variance. The amplitude of the random force, $\sqrt{2k_B
T\zeta}$, is chosen in accordance with the fluctuation-dissipation
theorem.

The equations of motion were integrated by a generalized form of the
velocity Verlet algorithm that has been recently proposed by Peters
\cite{peters04}.  For the time step of the integration we used $\delta
t=0.0083\,\tau$ (with the time unit $\tau=\sqrt{m d^2/\epsilon}$). First,
the samples were fully equilibrated in the temperature range $1.0 \ge T
\ge 0.14$, performing at least 30 independent runs at each temperature.
Starting from fully equilibrated samples, production runs at various
shear rates were performed. These runs were sufficiently long to reach 
the steady state regime.

At the lowest temperature $T=0.14$, 250 independent runs were done, each
of them over 40 million time steps. This relatively large effort was
necessary for an accurate determination of the shear stress, defined by
\begin{equation}
 \langle \sigma_{xy} \rangle =
  \frac{1}{L^3} \left\langle \sum_i 
   \left[ m_i \left( v_{i,x} - v_{{\rm s}, x}(y) \right) v_{i,y} 
         + \sum_{j>i} r_{ij,x} F_{ij, y} \right] \right\rangle \; .
\end{equation}
The shear stress $\langle \sigma_{xy} \rangle$ is a collective quantity
and thus lacks the self-averaging property of one-particle quantities.
Therefore, a relatively large number of independent runs have to be made
to obtain $\langle \sigma_{xy} \rangle$ with a reasonable accuracy. Below
we also present simulation results for the mean squared displacement
(MSD) of a tagged particle,
\begin{equation}
  \label{MSDeq}
  \langle r^2(t) \rangle =
     \langle \left( r_{{\rm tag},\alpha}(t) - r_{{\rm tag},\alpha}(0) 
             \right)^2 \rangle, 
     \quad \quad \alpha = x, y, z \ ,
\end{equation}
where $r_{{\rm tag},\alpha}(t)$ is the $\alpha$ component of the position
of the tagged particle at time $t$ (an average is performed over all
particles of the same species). The MSD is an example of a one-particle
quantity. From the MSD, the self-diffusion constant $D$ can be calculated
via the Einstein relation,
\begin{equation}
   D = \lim_{t\to \infty} \frac{\langle r^2(t) \rangle}{2t} \ .
  \label{eq_diff}
\end{equation}
Note that for the sheared system, the MSD as well as the self-diffusion
constants are anisotropic and depend on the considered Cartesian direction
(see below).

Another quantity, that we use to characterize the difference between
sheared and unsheared systems, is the so-called non-Gaussian parameter
\cite{nijboer66,boon},
\begin{equation}
  \alpha_2(t) = \frac{1 \langle r^4(t) \rangle}{3 \langle r^2(t) \rangle^2} -1 \; .
  \label{eq_alpha2}
\end{equation}
$\alpha_2(t)$ is the coefficient of first correction term to the Gaussian
approximation of the incoherent intermediate scattering function. Note
that $\alpha_2(t)$ vanishes in the diffusive long-time limit when the 
Einstein relation (\ref{eq_diff}) holds.

\subsection{Experimental techniques}
\subsubsection{Samples}

We used polymethylmethacrylate (PMMA) colloids fluorescently labeled with
nitrobenzoxadiazole (NBD) and dispersed in a mixture of cycloheptyl
bromide and cis-decalin that closely matches the density and refractive
index of the colloids. Since the colloids acquire a small
charge in this solvent mixture, we added $4\,\text{mM}$
of tetrabutylammonumchloride in order to screen the charges \cite{yethiray03}. Such a
system shows nearly hard-sphere (HS) behaviour, where the volume fraction
$\phi=(\pi/6) n d^3$ is the only thermodynamic control parameter.
 
The particle diameter $d = 1680 \pm 4\,\text{nm}$ was determined by
static light scattering on a very dilute colloidal suspension ($\phi\simeq
10^{-4}$). The diameter was also independently estimated from the position
of the first peak of the radial distribution function obtained by confocal
microscopy \cite{jenkins05,jenkins08}, yielding $d=1690\,\text{nm}$. With
the very dilute sample we also performed dynamic light scattering. This
confirmed the determined particle diameter and, in addition, allowed us
to deduce the relative polydispersity in size, $0.062$, from the angular
dependence of the diffusion coefficient \cite{pusey84}.

The colloid volume fraction of the stock solution was calibrated by
drying. A drop of the suspension was weighed and allowed to dry in a
vacuum oven until all the solvent was evaporated. The weight fraction was
then calculated as $\phi_w = m_{\rm dry}/m_{\rm total}$, where $m_{\rm
dry}$ is the mass of the dried sample and $m_{\rm total}$ the original
mass before drying. The colloid volume fraction was estimated from the
weight fraction as $\phi=\alpha\phi_w$, where the factor $\alpha$ takes
into account the contribution of the PHSA hairs, which are collapsed in
the dried state. Following \cite{pusey88} we assume $\alpha=1.04$. Our
sample was prepared at $\phi = 0.57$. The volume fraction of the sample
was also determined by direct imaging and found to be consistent.

\subsubsection{Shear Cell}
Shear was applied to the sample by means of a shear cell designed
and built in our lab. It represents an improved version of cells
previously used in light scattering and microscopy experiments
\cite{petekidis02a,petekidis02,smith07}. For the present microscopy
experiments, a specially designed microscope stage is used to mount
the shear cell in the microscope.  The cell consists of two parallel
glass coverslips (thickness $170\,\mu\text{m}$) whose separation can
be continuously varied within the range $300\ldots1000\,\mu\text{m}$
by vertically shifting the upper plate, which is sitting on three
metal spheres. Adjusting the position of the metal spheres also
ensures that the two plates are parallel;  this is verified as
described before \cite{smith04}. In the present experiment we used
a gap width ($320\,\mu\text{m}$) close to the minimum value in order
to achieve maximum strain. The bottom plate of the cell is driven by
a piezoelectric actuator (PI Instruments, model P-841.60) which can
provide a maximum stroke of $90\,\mu\text{m}$. The other end of the
bottom plate is connected to a pivoted lever driving the upper plate
in the direction opposite to the bottom plate \cite{petekidis02a}. The
vertical position of the pivot moves the vertical position of the
stationary plane inside the sample and changes the maximum displacement
of the top plate. In this experiment the maximum displacement of the
top plate was $630\,\mu\text{m}$, as measured by a displacement sensor
(Lion Precision, model ECL100-U8A, displacement range $2\,\text{mm}$)
which was calibrated prior to the measurement using the piezoelectric
actuator displacement, which is known with nanometer accuracy. With this
arrangement the maximum obtainable strain is $\gamma=2.25$.

A constant shear rate is achieved by driving the piezoactuator with a
linear voltage ramp which covers the whole displacement range.  The slope
of the ramp determines the shear rate $\dot{\gamma}$. The shear rate is
also experimentally measured from the velocity profile across the gap,
as explained below.

In order to minimize solvent evaporation during the experiments, the
two glass plates are surrounded by teflon sheets which touch each other
and thus seal the sample environment. Due to the very low friction of
teflon, no change in the plate displacement due to the contact of the
teflon sheets is observed, as verified during calibration.

In recent investigations of the flow of glassy colloidal dispersions wall
slip was observed when smooth surfaces were used to apply shear to the
samples. It was shown \cite{Besseling2007} that slip can be prevented
when the surfaces are coated with colloidal particles similar to those
used for sample preparation.  Although our volume fraction is below the
glass transition, we coated the coverslips with a layer of polydisperse
colloids. The velocity profile across the gap indeed shows no indication
of wall slip (see below).

\subsubsection{Confocal Microscopy and Image Analysis}
Confocal Microscopy experiments were performed with a fast-scanning
VT-Eye confocal microscope (Visitech International) mounted on a
Nikon TE2000-U inverted microscope. A Nikon Plan Apo VC 100$\times$
oil immersion objective was used for all measurements. Two-dimensional
images of the samples were recorded at a depth of $15\,\text{$\mu$m}$
inside the sample (except for the determination of the velocity profile,
see below), in order to avoid boundary effects and to retain good quality
images. Images have $512\times512$ pixels, corresponding to an area of
$57\times57\,\text{$\mu$m}^2$.

In a typical experiment 500 images were recorded starting simultaneously
with the application of shear and ending when the maximum strain was
reached. Image acquisition and application of shear were synchronized by
starting the voltage ramp, which controls the displacement of the plates,
with a trigger provided by the confocal microscope. The response of the
piezoactuator to the voltage ($\text{$\mu$s}$ delay) can be considered
instantaneous on the time scale of our experiments.

\begin{figure}
\begin{center}
\includegraphics[width=.55\textwidth]{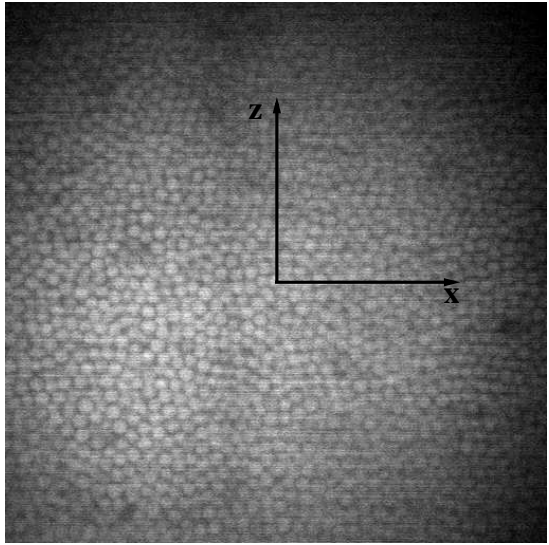}
\caption{Confocal microscopy image of the sample at switch on of shear. 
The axes $x$ and $z$ indicate respectively the velocity and vorticity 
direction. The gradient direction is perpendicular to the plane of the 
image. \label{confocal_image}}
\end{center}
\end{figure}

\begin{figure}
\begin{center}
\includegraphics[height=.55\textwidth,angle=270]{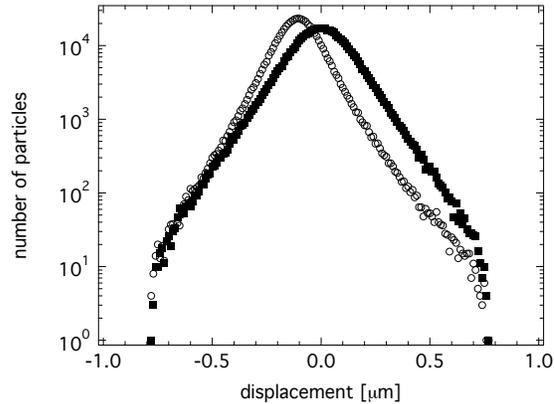}
\caption{Histogram of particle displacements between consecutive images along
the velocity $x$ ($\circ$) and vorticity $z$ ($\blacksquare$) direction.
\label{disp_hystogram}}
\end{center}
\end{figure}
Figure \ref{confocal_image} shows a typical confocal image acquired in
experiments.  The axes $x$ and $z$ indicate the velocity and vorticity
direction. The gradient direction, $y$, is perpendicular to the plane of
the image. Particle coordinates and particle trajectories were extracted
from images using standard routines \cite{crocker96}. The shear rate
applied here ($\dot{\gamma} =  0.045\,\text{s}^{-1}$) is sufficiently
small so that particles do not move very far between two consecutive
frames even in the $x$ direction. It is thus not necessary to remove
the affine motions before tracking \cite{isa06}. This is supported by the
histograms of the frame by frame displacements in $x$ and $z$ directions
as determined during a step rate experiment (Fig.~\ref{disp_hystogram});
both histograms lie well within the displacement axis, which covers the
maximum detectable range (about $\pm 1\,\mu$m).

Particle trajectories were used to calculate mean-squared displacements
(MSD) and non-Gaussian parameters ($\alpha_2$) in the vorticity
direction $z$, according to equations \ref{MSDeq} and \ref{eq_alpha2}
respectively. Only trajectories starting at commencement of shear are
included in the MSD and $\alpha_2$ calculations in order not to loose
information on the time dependence of the MSD and $\alpha_2$ after
switching on the shear field. This implies that in each experiment only
a limited number of particles are contributing to the average value
of the two quantities; initially of the order of 1000 and the number
decreasing with time as particles leave the field of view due to motion
in or out of the observation plane. In order to improve statistics,
MSDs and $\alpha_2$ extracted from several, typically 20, experiments
were averaged after checking reproducibility.

The velocity profile across the gap was determined immediately after the
application of shear by recording image series at different depths into
the sample. The depth was increased in steps of $10\,\text{$\mu$m}$, from
the first layer of particles above the coating up to $100\,\text{$\mu$m}$
into the sample. Deeper in the sample image quality did not allow for
a quantitative analysis. Displacements between consecutive images were
determined by maximizing the correlation between consecutive images
\cite{isa06}. The obtained velocity profile (Fig.~\ref{fig5}) shows a
linear dependence with a shear rate $\dot{\gamma}=0.0458\,\text{s}^{-1}$,
in good agreement with the nominal value (for a detailed discussion
see below).

\subsection{Mode coupling theory for glassforming liquids under shear}
Recent advances \cite{fuchs02,fuchs05,brader07,brader08pre} have
generalized the mode coupling theory of the glass transition to describe
dense colloidal suspensions in a flowing solvent, characterized by a
non-vanishing velocity gradient tensor of arbitrary time dependence
$\bs\kappa(t)$.  The resulting non-linear theory provides a route
to calculating non-equilibrium averages of functions of the particle
coordinates.  In particular, a closed microscopic expression for the
shear-stress in terms of the full flow history of the system, i.e.,
a constitutive equation relating stress and strain, may be obtained.
The starting point of the theory is the many-body Smoluchowski equation
\cite{dhont}
\begin{equation}\label{smoluchowski}
  \partial_t\Psi(t)=\Omega(t)\Psi(t)
  \equiv\sum_i\bs\partial_i\cdot\left[D_0(\bs\partial_i
  -\beta\bs F_i)-\bs\kappa(t)\cdot\bs r_i\right]\Psi(t)\,,  
\end{equation}
where $D_0$ is the bare diffusion coefficient and $\beta=1/k_BT$.
Equation (\ref{smoluchowski}) describes the evolution of the probability
distribution function of particle positions $\Psi(t)$ in the overdamped
limit.  This formulation of the theory omits hydrodynamic interactions and
assumes a prescribed, spatially constant $\bs\kappa(t)$, thus excluding
confinement effects and inhomogeneous flows  such as shear-banded and
shear-localized states.  Although the assumption of translationally
invariant flow is clearly an idealization, the simulation and experimental
results presented in this work suggest that Eq.(\ref{smoluchowski})
is a reasonable starting point, at least for dense fluid states under
shear flow.  In the case of a constant shear flow $\dot\gamma$ along the
$x$-direction and with gradient in the $y$-direction the velocity gradient
tensor is given by $\kappa_{ij}=\dot\gamma\,\delta_{ix}\delta_{jy}$.

The integration through transients formalism
\cite{fuchs02,fuchs05,brader07,brader08pre} provides a formal solution
to Eq.(\ref{smoluchowski}) by integrating over the entire flow history.
In the case of shear, the following formal result is obtained for the
time-dependent distribution function
\begin{equation}\label{itt}
\Psi(t)=\Psi_e + \int_{-\infty}^tdt'\,
   \dot\gamma(t')\; \Psi_e\hat\sigma_{xy}
   {\rm e}_{-}^{\int_{t'}^t ds\,\Omega^\dagger(s) }\,,
\end{equation}
where $\hat\sigma_{xy}=-\sum_iF_{ix}r_{iy}$ is the potential part of the
microscopic stress tensor and $\Omega^\dagger$ is the adjoint Smoluchowski
operator. The time-ordered exponential function ${\rm e}_{-}$ arises 
because $\Omega^\dagger(t)$ does not commute with itself for different times.
The assumption of an equilibrium Boltzmann distribution $\Psi_e$ in the
infinite past allows general nonequilibrium averages to be expressed
in terms of averages taken with the equilibrium distribution function.
Equation~\eqref{itt} is an operator expression, to be used with the
understanding that quantities to be averaged are placed to the right
before integration over particle coordinates.  Within linear response
the macroscopic shear stress $\sigma(t)$ is given in terms of the shear
modulus by the familiar Green-Kubo formula of equilibrium statistical
mechanics.  Using (\ref{itt}) to calculate the average of the potential
part of the macroscopic stress, viz. $\hat\sigma_{xy}/V$, where $V$ is
the system volume, yields a non-linear generalized Green-Kubo relation
\cite{fuchs02,fuchs05,brader07,brader08pre}
\begin{equation}\label{stressgeneral}
\sigma_{xy}(t)=\frac{1}{V}\int_{-\infty}^t\dot\gamma(t')
\langle\hat\sigma_{xy}{\rm e}_{-}^{\int_{t'}^t ds\,\Omega^\dagger(s) }
\hat\sigma_{xy}\rangle,
\end{equation}
where $\langle\cdot\rangle$ represents an average over the equilibrium
distribution. Note that the flow history $\dot\gamma(t')$ appears both
explicitly in Eq.~\eqref{stressgeneral} (recovering linear response) and
implicitly (nonlinearly) through the time-evolution of the stress-stress
correlation function.  In the present situation of start-up shear,
$\dot\gamma(t')$ vanishes for all $t'<0$, so that the integration
is only performed for $t'>0$, where the Smoluchowski operator has no
explicit time dependence.  Defining the generalized dynamical shear
modulus by $G(t)=\langle\hat\sigma_{xy}(t)\hat\sigma_{xy}(0)\rangle/V$
as a non-linear function of $\dot\gamma$, we have
\begin{equation}
\sigma_{xy}(t)= \dot\gamma\int_0^{t}\!\! dt' G(t')\,.
\label{stress}
\end{equation}
Within the mode-coupling approach $G(t)$ is approximated by projecting
the dynamics onto density-pair modes corresponding to all possible
wave-vector pairs and directions.  The considerable numerical complexity
of the resulting equation can be significantly reduced by employing the
following isotropic approximation for the modulus \cite{faraday},
\begin{eqnarray}
G(t)=\frac{k_BT}{60\pi^2}\int\! dk\,
\frac{k^5}{k(t)} \frac{S'_k S'_{k(t)}}{S^2_{k}}\Phi^2_{k(t)}(t)\,,
\label{modulus}
\end{eqnarray} 
where $S_k$ is the static structure factor.  The affine solvent flow
enters via the time-dependent wavevector $k(t)$ describing the advection
of density fluctuations for wave number $k$ to smaller wavelengths. The
exact anisotropic advection $k(t)=\sqrt{k^2 + 2k_xk_y\dot\gamma t
+ k_x^2\dot\gamma^2 t^2}$ needs to be approximated by an isotropic
wavevector in the mode-coupling $k$-integrals, $k(t) \approx k\sqrt{1 +
(\dot\gamma t /\gamma_c)^2/3 }$, \rem{while reducing substantially} in
order to reduce the computational resources required.  Pre-averaging
over spatial directions introduces an additional source of error
when compared to full solution of the anisotropic MCT equations.
To first order, we expect, this discrepancy can be compensated by
introducing a characteristic strain parameter, $\gamma_c$, to which
we assign the value $\gamma_c=0.1$.  Evaluation of (\ref{modulus})
requires the transient intermediate density correlator, defined by
$\Phi_k(t) =\langle\varrho_{\bs k}^*\exp[\Omega^\dagger t]\varrho_{\bs
k(-t)}\rangle/NS_k$.  The appearance of an advected wavevector in this
definition removes the trivial decorrelation of density fluctuations
arising from purely affine flow.  As a consequence of the equilibrium
averaging, $\Phi_k(t)$ only contains information regarding the strain
accumulated between $t=0$ and later time $t$.  Mori-Zwanzig type
projection operator manipulations yield an equation of motion for the
correlator \cite{faraday}
\begin{eqnarray}
\frac{\partial}{\partial t}\Phi(t) + \Gamma_q\left(
\Phi_q(t) + \int_0^{t}\!dt'\, m_q(t-t')\frac{\partial}{\partial t'}\Phi_q(t')
\right)=0,
\label{eom}
\end{eqnarray}
where $\Gamma_q=q^2D_0/S_q$ is the initial decay rate.  Mode-coupling
approximations provide an explicit form for the memory function
\begin{eqnarray}
\hspace*{-1cm}
m_q(t)=\int \!d{\bf k}\, \frac{n S_q S_k S_p}{16\pi^3q^4}
[\,{\bf q}\cdot{\bf k}\,c_{k(t)} + {\bf q}\cdot{\bf p}\,c_{p(t)}\,]\,
[\,{\bf q}\cdot{\bf k}\,c_{k} + {\bf q}\cdot{\bf p}\,c_{p}\,],
\label{memory}
\end{eqnarray}
where $n$ is the number density and $c_q$ is the equilibrium direct
correlation function, $n c_q=1 - 1/S_q$.  It should be noted that the
colloid-colloid interaction potential and thermodynamic statepoint enter
purely via the equilbrium structure factor.  In the absence of flow the
two vertices in (\ref{modulus}) or  (\ref{memory}) form a perfect square.
For finite shear rates wavevector advection leads to a de-phasing of
the vertices, resulting in a reduction of memory effects. This is the
mechanism by which the competition between slow structural relaxation
and flow enter the theory.  Equations \eqref{modulus}-\eqref{memory}
form a closed theory for the calculation of $G(t)$. The corresponding
shear stress follows from \eqref{stress}.

For the purposes of this study we consider a one-component system of
hard-spheres of diameter $d$, and all numerical calculations will be
performed using the Percus-Yevick approximation for $S_q$; discretization
of the integrals is performed like in Refs. \cite{fuchsmayr,faraday}.
This minimal model provides both a reasonable first approximation to
the PMMA colloids used in our experiments and captures the fundamental
excluded volume effects responsible for the phenomenology observed in our
binary mixture simulations.  The relevant parameters in the hard-sphere
system are the bare P\'eclet number ${\rm Pe}_0=\dot\gamma d^2/D_0$,
measuring the importance of shear relative to Brownian motion, and the
packing fraction $\phi$ which determines the timescale of structural
relaxation in equilibrium. The effect of shear on the structural
relaxation is measured by the dressed P\'eclet or Weissenberg number Pe,
which we define by ${\rm Pe} = \dot\gamma d^2 / ( 2 D )$, using $D$,
the long time self diffusion coefficient in equilibrium. It compares the
shear rate with the time required for a particle to diffuse  a distance
of its diameter in one spatial direction.

\section{Results}\label{sec:results}
\begin{figure}
\begin{center}
\includegraphics[width=0.55\textwidth]{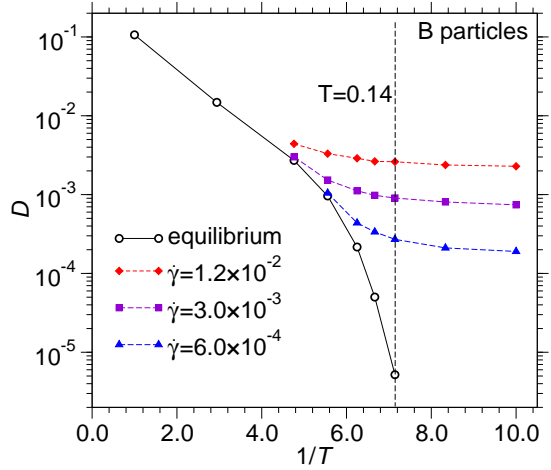}
\caption{\label{fig1}
Self-diffusion constants for the B particles in equilibrium and at
the indicated shear rates, as obtained from the simulation. For the
sheared systems the diffusion coefficients are calculated from the mean
squared displacement in $x$ and $y$ direction, i.e.~perpendicular to
the shear direction. The temperature $T=0.14$ is marked.}
\end{center}
\end{figure}
Figure~\ref{fig1} shows the diffusion coefficients $D(\dot\gamma,T)$
obtained from the simulations of the binary Yukawa mixture at various
temperatures and shear rates.  Only the diffusivity of the larger
particles is shown, as the smaller particles behave qualitatively
identically. Open circles represent values measured in the quiescent
equilibrium state, $D(\dot\gamma=0, T)$. They show a strong
faster-than-Arrhenius slowing down with decreasing temperature, signifying
that the system approaches a glass transition. On the timescale of the
simulation, no diffusive regime is reached in the MSD for temperatures
lower than $T=0.14$.

Applying steady shear to the system, one finds instead that even at
the lowest temperatures considered here, the system shows a finite
diffusivity. For a fixed value of $\dot\gamma$, the $D(\dot\gamma, T)$
curves approach a plateau as $T$ is lowered.  This plateau occurs at
values increasingly larger than the quiescent diffusion coefficient. Such
an effect is a manifestation that the system shows shear-thinning
behaviour.  At high temperatures, a finite $\dot\gamma$ has only a small
effect on the diffusion coefficient, and consequently all curves shown
in the figure approach each other for small $1/T$.

To study the effects of switching on shear flow suddenly, we consider
a fixed temperature $T=0.14$, corresponding to the lowest temperature
where we could equilibrate the quiescent system.  This allows us to
start from a well-defined reference state at $t=0$, and enables direct
comparison with the theory. It also precludes aging effects arising from
the unsheared state.

\begin{figure}
\begin{center}
\includegraphics[width=0.55\textwidth]{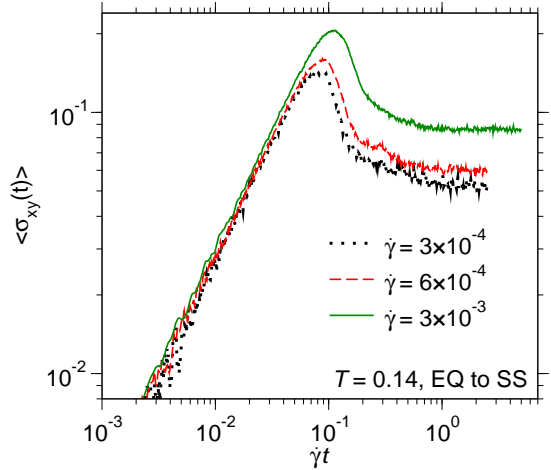}
\caption{\label{fig2}
Stress-strain relation from the simulation at the temperature $T=0.14$
for three different shear rates, as indicated.
}
\end{center}
\end{figure}
The macroscopic stress $\sigma(t)$ measured from the time of switching
on steady shear is shown in Fig.~\ref{fig2} for three different shear
rates. The kinetic contribution is negligible compared to the potential
one, the one studied in the theory. All three curves are qualitatively
similar, and show three regimes. At early times, corresponding to small
strains $\gamma=\dot\gamma t\lesssim0.1$, $\sigma(\gamma)$ increases
almost linearly with $\gamma$.  This is the regime of solid-like elastic
response, and from the prefactor of this almost linear increase one
can infer the elastic constant. It depends weakly on $\dot\gamma$.
At large strains, $\gamma\gtrsim1$, one finds on the other hand that
$\sigma(t)$ approaches an asymptotic value attained in steady state,
the dynamic shear stress $\sigma(\infty)$ resulting from keeping the
system flowing at a fixed rate. This corresponds to the regime of plastic deformation.
The long-time plateau increases with increasing $\dot\gamma$, indicating
that faster flow induces higher internal stresses.
The dependence of $\sigma(\infty)$ on shear rate is called 'flow curve'.

At strains in the intermediate range, $0.1\lesssim\gamma\lesssim1$,
a marked overshoot is seen in $\sigma(\gamma)$: the almost linear
increase for small strains continues beyond the steady-state value of
$\sigma$, so that a maximum $\sigma_\text{max}>\sigma(\infty)$ results
at roughly $10$\% strain in units of the particle diameters, slightly
increasing with increasing shear rate. Only after this maximum does
$\sigma(\gamma)$ decay towards its steady-state value. Similar features
have been observed in experiment \cite{Osaki2000,Islam2001,overshoot}
and simulation \cite{heyes86,Varnik2004,Rottler2003,tanguy06}.

\begin{figure}
\begin{center}
\includegraphics[width=0.55\textwidth]{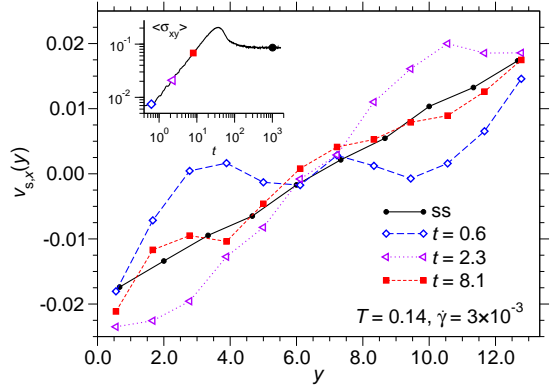}
\caption{\label{fig3} Evolution of the shear profile from equilibrium 
to steady state, as obtained from the simulation of the binary Yukawa 
mixture. Between the equilibrium and the steady state
profiles are shown at the times $t=0.6$, $t=2.3$ and $t=8.1$.
The applied shear rate is $\dot{\gamma}=3\times10^{-3}$. The temperature
is $T=0.14$. The inset shows the stress as a function of time for 
$\dot{\gamma}=3\times10^{-3}$. The symbols in the inset indicate the
times along the stress curve at which the shear profiles are shown.}
\end{center}
\end{figure}
Since the simulation introduces shear flow only through the boundary
conditions, the sudden switching on of shear does not induce a linear
flow profile instantaneously throughout the simulation box. Rather,
there exists some time scale over which the steady-state velocity
profile builds up. Especially since we intend to compare the simulation
results with MCT, where a fixed velocity profile is presumed to exist
always, let us point out that none of the prominent features shown in
Fig.~\ref{fig2} depend on this gradual self-adjustment of the profile.
To this end, we determined the instantaneous velocity profiles, i.e.,
$v_{{\rm s}, x}(y)$, for the $\dot\gamma=3\times10^{-3}$ simulation
at various times following shear startup.  They have been obtained
by averaging over a negligibly small time window and over the 250
independent simulation runs. As Fig.~\ref{fig3} shows, the profile
for $t=0.6$ indeed still shows a pronounced `S'-shaped form, as the
Lees-Edwards boundary conditions essentially introduce small sheared
boundary zones to the unperturbed bulk. The propagation of these shear
zones inwards is, however, fast. Already around $t=8.1$, corresponding
to a strain of $\gamma\approx0.02$, one essentially recovers a linear
velocity profile across the whole system within the error bars. This
observation also holds for larger times (not shown). For comparison, also
the profile obtained in the steady-state part of the simulation (averaged
over a longer time interval and hence displaying smaller fluctuations)
is shown in the figure. We do not find the formation of shear bands or
similar localized features in the flow profile, even though the simulated
equations of motion do not bias towards a linear profile by construction.

Comparing the strain at which the steady-state flow profile is first
attained with the typical strains identified from the overshoot features
in Fig.~\ref{fig2}, we conclude that the latter are not connected to the
peculiarities of whether the system is sheared only at the boundaries
or forced to follow a linear profile mediated through the solvent
immediately. This makes comparisons to MCT viable.

\begin{figure}
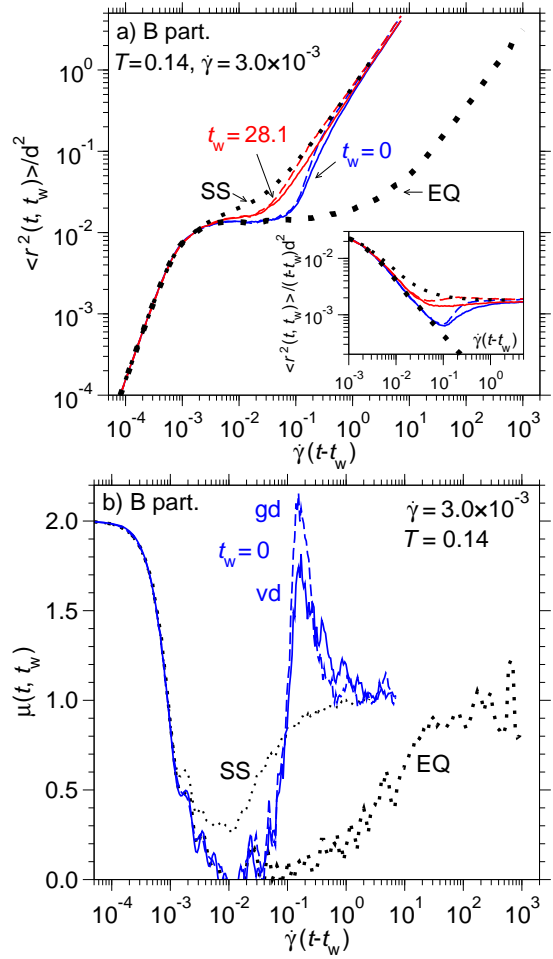

\begin{center}
\includegraphics[width=0.55\textwidth]{fig6a.eps}\\
\includegraphics[width=0.55\textwidth]{fig6b.eps}
\caption{\label{fig4}
a) Log-log plot of the mean squared displacements of B particles at the
temperature $T=0.14$ as a function of strain $\gamma\equiv\dot{\gamma}t$.  Dashed and
dotted lines correspond to equilibrium and steady state rescaled with $\dot{\gamma}
= 3\times 10^{-3}$.  The solid and dashed lines show the transition from
equilibrium to steady state at this $\dot\gamma$ for the vorticity direction ($z$, 'vd') and the gradient
direction ($y$, 'gd'), respectively. Data for the waiting times $t_{\rm w}=0$
and 28.1 are shown (from right to left). The dotted lines correspond
to the mean squared displacements in the equilibrium (EQ) and in the
steady state (SS), as indicated.  In the inset, $\langle r^2(t,t_{\rm
w}) \rangle / (t-t_{\rm w})$ is plotted as a function of strain.  b)
Effective exponent $\mu(t)=d[\log\langle r^2(t,t_{\rm
w}) \rangle]/d[\log (t-t_{\rm w})]$ for some of the curves shown in a).}
\end{center}
\end{figure}
More information on the transient dynamics of the system can be
obtained from two-time correlation functions, formed between two
times $t_{\rm w}>0$ and $t>t_{\rm w}$. As a simple example, we show in
Fig.~\ref{fig4} the mean-squared displacements (MSD) of B particles,
for three different waiting times, $t_{\rm w}=0$, $t_{\rm w}=28.1$,
and $t_{\rm w}$ sufficiently large so that the result as a function of
$t-t_{\rm w}$ becomes $t_{\rm w}$-independent.  For comparison, also
the equilibrium MSD is shown. It agrees with the MSDs under shear for
short times.  Note that the solid lines correspond to the MSD evaluated
in the gradient direction, while the dotted equilibrium and steady-state
lines represent averages over both directions perpendicular to shear.
Comparing steady-state and equilibrium, we recover for long times
the shear-thinning effect discussed in connection with the diffusion
coefficients, Fig.~\ref{fig1}. The equilibrium curve shows a pronounced
plateau where $\langle r^2(t,t_{\rm w})\rangle\approx0.013=r^2_{\rm
plat}$, indicating that particles are at intermediate times caged,
with a localization length around  $r^{\rm l}\equiv r^2_{\rm plat}/6
\approx0.02$. This is the usual glassy dynamics described by MCT in
quiescent systems.  The steady-state curve does not show such a pronounced
plateau, due to the speed-up of the final relaxation.

As Fig.~\ref{fig4} shows, the sudden start-up of shear flow has a drastic
effect on the MSD measured immediately at the start-up time. While
the curve follows the quiescent MSD for strains $\gamma\lesssim0.1$,
for larger time it suddenly starts to increase much more rapidly,
and coincides with the steady-state MSD already for $\gamma\approx1$,
although $D(\dot\gamma)\gg D(\dot\gamma=0)$.  This effect is seen
in both directions perpendicular to the flow, albeit somewhat less
pronounced in the vorticity direction, as demonstrated by the solid
lines in Fig.~\ref{fig4}.  There is thus an intermediate time window
where the MSD shows super-diffusive behaviour, i.e., grows faster than
$t$. This is even more clearly seen when plotting $\langle r^2(t,t_{\rm
w})\rangle/(t-t_{\rm w})$, as shown in the inset of Fig.~\ref{fig4}.
At long times, this curve would monotonically fall to $2D$ in the
quiescent system. A dip followed by an increase at intermediate times,
as seen in our data, is the signature of superdiffusion.

The effect can be quantified more precisely through a logarithmic
derivative, $\mu(t)=d[\log\langle r^2(t,t_{\rm w})\rangle]/d[\log
(t-t_{\rm w})]$, which approaches $\mu=1$ for ordinary diffusion
and $\mu=2$ for ballistic motion. The simulation result for $\mu(t)$
and $t_{\rm w}=0$ is shown both for the vorticity direction and the
gradient direction in the lower panel of Fig.~\ref{fig4}. At short times,
$\mu(t\to0)\to2$ reveals the ballistic motion underlying the molecular
dynamics simulation. At $t\approx1$, $\mu(t)\approx0$ is indicative of the
equilibrium cage effect. In the $t_{\rm w}=0$ curve, it is truncated at
$\dot\gamma t\approx0.1$, and quickly $\mu(t)$ raises to about $2$ in the
gradient direction, and $1.8$ in the vorticity direction, before finally
settling to $\mu(t\to\infty)\to1$ as expected.  The simulation thus
reveals an almost `ballistic' regime in the MSD at intermediate times.
Remarkably, this result remains unchanged if one performs the simulation
with a much larger damping coefficient in the DPD thermostat, $\xi=1200$
\cite{zauschpre}. This highlights that the regime of superdiffusion is
likely a feature of sheared glassy dynamics and not of the short-time
microscopic motion.  It is worth noting that $\mu\le1$ holds for all
times beyond the microscopic transient in the steady-state curve. This
is what one expects from the regime of true `structural relaxation',
and what holds precisely in the Brownian equilibrium system, where all
correlation functions are purely relaxing functions.

\begin{figure}
\vspace*{0.2cm}
\begin{center}
\includegraphics[width=0.55\textwidth]{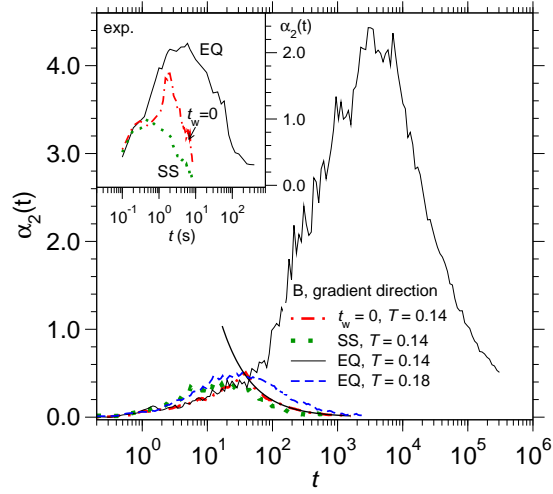}
\caption{\label{alpha2_gd} Non-Gaussian parameter $\alpha_2(t)$
for equilibrium, steady state and waiting time $t_{\rm w}=0$ at
the temperature $T=0.14$. For the sheared case, the shear rate
is $\dot{\gamma}=3\times 10^{-3}$. For comparison also the equilibrium
$\alpha_2(t)$ at $T=0.18$ is shown. $D(\dot{\gamma}=0)$ at this
temperature is approximately equal to $D(\dot{\gamma}=3\times 10^{-3})$
at $T=0.14$. The bold solid line is a fit to the $t_{\rm w}=0$ curve
with the indicated power law $\alpha_2 \propto t^{-0.95}$. The inset shows
experimental results for $\alpha_2(t)$. Here, the shear rate is
$\dot{\gamma}=4.5\times10^{-2}$\,s$^{-1}$.}
\end{center}
\end{figure}
Figure~\ref{alpha2_gd} displays the non-Gaussian parameter $\alpha_2(t)$,
as defined by Eq.~(\ref{eq_alpha2}), for equilibrium, steady state
and waiting time $t_{\rm w}=0$ at $T=0.14$ and $\dot{\gamma} = 3\times
10^{-3}$. Also shown is the equilibrium $\alpha_2(t)$ at $T=0.18$ where
$D(\dot{\gamma}=0)$ is approximately equal to $D(\dot{\gamma}=3\times
10^{-3})$ at $T=0.14$.  In the steady state the amplitude of maximum
in $\alpha_2(t)$ is about an order of magnitude smaller than that for
the equilibrium at $T=0.14$ and also slightly smaller than that of the
equilibrium curve at $T=0.18$.  Thus the application of shear seems
to reduce non-Gaussian effects at intermediate times. Similar to the
behaviour of the MSD, the non-Gaussian parameter for waiting time $t_{\rm
w}=0$ follows first the equilibrium curve. Then, at time $t \gtrsim 30$
the decay of $\alpha_2$ can be fitted to a power law,
$\alpha_2 \propto t^{-0.95}$.

\begin{figure}
\begin{center}
\includegraphics[width=0.55\textwidth]{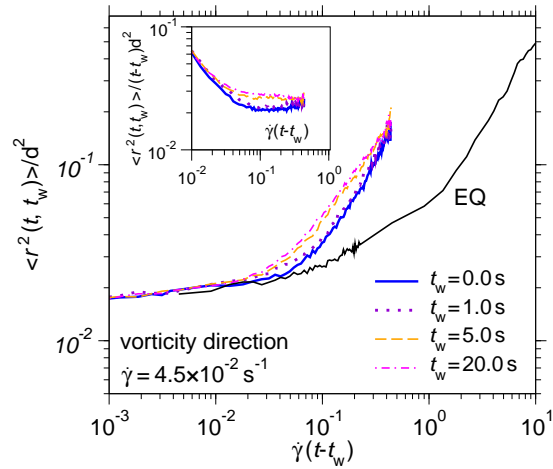}
\caption{\label{fig6} Log-log plot of the experimental mean squared 
displacement in vorticity direction as a function of 
strain $\dot{\gamma}t$ for different waiting times, as indicated. The inset shows 
$\langle r^2(t,t_{\rm w}) \rangle / (t-t_{\rm w})$
as a function of $\dot{\gamma}t$.}
\end{center}
\end{figure}
Let us now compare in detail the findings from the simulation to
the colloidal experiment. Figure~\ref{fig6} shows the mean-squared
displacements obtained from confocal microscopy and particle tracking,
taken in the vorticity direction. Here only a single shear rate
is shown. Note in comparing with Fig.~\ref{fig4}, that the relevant
quantities characterizing the magnitude of shear differ between the two
setups: while the experiment corresponds to a Weissenberg number ${\rm
\We}\approx 7$ (Pe$_0 \approx 1$), it is closer to ${\rm \We}\approx 288$
in the simulated system.  Apart from obvious differences due to this
circumstance, the results are qualitatively similar.  As the inset of
Fig.~\ref{fig6} demonstrates for the experimental $\langle r^2(t,t_{\rm
w})\rangle/(t-t_{\rm w})$, again a nonmonotonic dip is found at roughly
$\gamma\approx0.1$, identifying a corresponding super-diffusive regime
in the MSD. Like in the simulation, this effect is most pronounced
for $t_{\rm w}=0$ and then continuously weakens as the steady state
is approached.

A quantitative difference is found when comparing the extension and
strength of the super-diffusive window in the MSD. In the simulation,
the crossover in the $t_{\rm w}=0$ curve from the equilibrium to
the steady-state limiting cases is much more rapid. Likewise, the
experiments show the first increase of the transient MSD beyond the
equilibrium curve for slightly smaller strains, $\gamma\approx0.05$,
than the $\gamma\approx0.1$ found in the simulation.

\begin{figure}
\begin{center}
\includegraphics[width=0.55\textwidth]{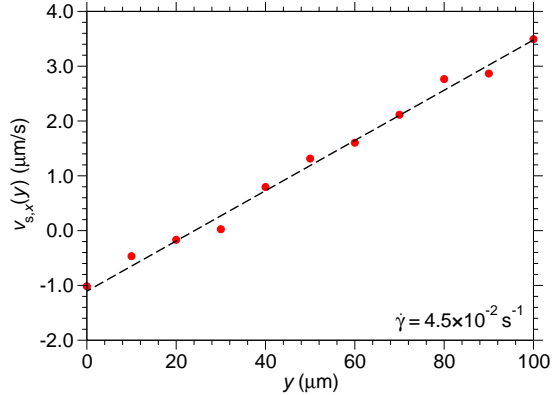}
\caption{\label{fig5}
Steady state shear profile, as obtained from the experiment (dots).
Dashed line: linear fit providing $\dot{\gamma}=0.045$\,s$^{-1}$.}
\end{center}
\end{figure}
Also the non-Gaussian parameters have been determined in the experiment.
They are shown in the inset of Fig.~\ref{alpha2_gd}.
One can infer that also this quantity behaves qualitatively
as in the simulation.

Again it is important to stress that the effects discussed here are not
consequences of shear localization or, more generally, developments
of non-linear flow profiles. This has also been checked for the
experiment. The velocity profile in steady state is almost perfectly
linear, as Fig.~\ref{fig5} demonstrates. This confirms that also in
the experimental setup, a homogeneous constant-shear steady state is
approached, without obvious shear-localization features or wall effects.

Having established the close connection between simulation and experiment,
let us now try to rationalize these findings with MCT.  The quantity
$\sigma(t)$ is easily accessed in the mode-coupling framework and thus
forms a convenient starting point for the analysis.  In order to provide a
meaningful comparison with simulation results, we have to first map the
relevant quantities between the theoretical hard-sphere model and the
binary Yukawa mixture employed in the simulations. Since we are dealing
with slow structural dynamics, it is natural to fix the\rem{ ratio of}
final\rem{ ($\tau$) to microscopic ($t_0$)} relaxation time $\tau$
first for equilibrium\rem{ (quantifying the slowness of equilibrium
diffusion through the ratio $\tau/t_0$, the `Deborah' number)}
and then to consider the effect of shear identifying the P\'eclet
and Weissenberg numbers. The only free parameter in the quiescent
hard-sphere model is the packing fraction, adjusted in the following
to $\Delta\phi=\phi-\phi_c=-1.16\times10^{-3}$ from the requirement
that MCT describes the equilibrium MSD found in the simulation for
long times (see below).  Next, the shear rate appropriate for use in
the theory may be determined.  We find ${\We}=183$ (corresponding to
$\dot\gamma=5.5\times10^{-3} D_0/ d^2$) \rem{$\dot\gamma=5.5\times10^{-3}
d^2/D_0$} to lead to the same $D(\dot\gamma)/D(\dot\gamma=0)$ as it
arises from the simulation at ${\We}\approx 288$.

\begin{figure}
\begin{center}
\includegraphics[width=.65\textwidth]{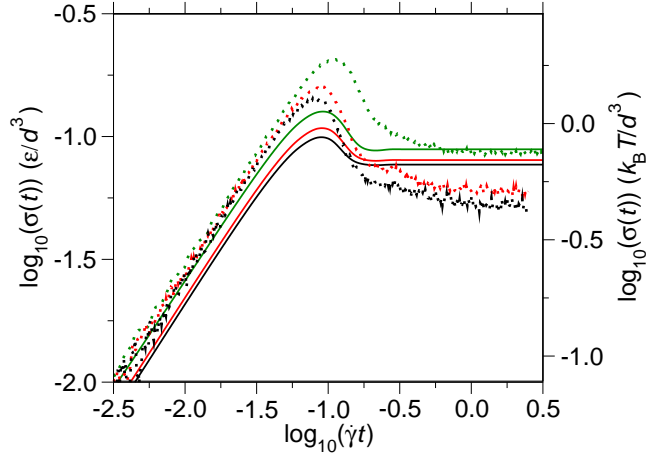}
\caption{
Comparison of the stress from simulation (dots) and theory (solid
lines) for the shear rates considered in simulation,
$\dot\gamma=3\times 10^{-3}$ (green), $\dot\gamma=6\times 10^{-4}$ (red)
and $\dot\gamma=3\times 10^{-4}$ (black). The relation between simulation
energy units ($E_{\rm md}$) and theory is $E_{\rm md}=8.8k_BT$.
The theoretical shear rates (in units of $d^2/D_0$) obtained from
fitting the MSD at the highest shear rates of the simulation
(corresponding to a theoretical value Pe=183) are
$\dot\gamma=5.5\times 10^{-3}$, $\dot\gamma=1.1\times 10^{-3}$ and
$\dot\gamma=5.5\times 10^{-4}$, respectively.
\label{joe1b}}
\end{center}
\end{figure}
In Fig.~\ref{joe1b}, we show the resulting MCT-calculated stress
$\sigma(t)$ as a function of accumulated strain $\dot\gamma t$ for the
three shear rates considered in the simulation.  Theoretical results are
shown in hard-sphere units ($k_BT/d^3$) whereas simulation results are
expressed in units of $E_{\rm md}=\epsilon/d^{3}$.  The two axes
have been adjusted in order to match the stress attained in steady state,
fixing a relation of $E_\text{md}\approx 8.8k_BT$ between the kinetic
energies of the different systems for the largest shear rate considered
in the simulation. This fitted value of $T$ is surprisingly close to the 
actual value in the simulation, indicating that the theory captures the stress magnitude well.
As Fig.~\ref{joe1b} shows, this also provides a
reasonable description of the initial small-strain stress, reflecting the
elastic constant of the system. More importantly, the theory qualitatively
reproduces the stress overshoot found in the simulation.  That the
peak occurs at strains of roughly $\gamma\approx0.1$, in accord with
simulation, is a consequence of our choice for the parameter $\gamma_c$
in the isotropic approximation to the advected wavevector.  The magnitude
of the overshoot is significantly underestimated, which might be due to
the MCT approximation per se, but may also be a result of the additional
isotropic approximation underlying our calculations.  In the decay of
$\sigma(t)$ after its maximum, the simulation data show a rather slow
decay for $\log_{10}\gamma\gtrsim0.75$, which is also not captured in MCT.

\begin{figure}
\begin{center}
\includegraphics[width=.55\textwidth]{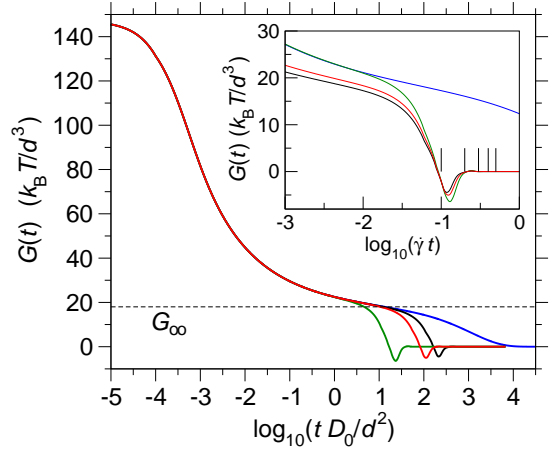}
\caption{Generalized shear modulus from mode-coupling theory,
calculated using $\phi-\phi_c=-1.16\times10^{-3}$, for $\dot\gamma=0$
(rightmost curve) and $\dot\gamma=5.5\times10^{-3}$, $1.1\times10^{-3}$,
and $5.5\times10^{-4}$ (left to right), corresponding to curves in
figure~\ref{joe1b}. The stress overshoot in Fig.~\ref{joe1b} is related
to the negative undershoot in $G(t)$ at long times.  The plateau value
which would develop closer to the glass transition is indicated by the
broken line.  The inset shows the moduli as a function of strain, where
the vertical bars indicate the strains considered in Fig.~\ref{structure}.
\label{joe2b}}
\end{center}
\end{figure}
The theory quite naturally relates this stress overshoot to a peculiar
feature of the dynamic shear modulus $G(t)$. In Fig.~\ref{joe2b},
this modulus is shown both in the quiescent case and for the shear
rates used in Fig.~\ref{joe1b}.  The equilibrium function shows the
two-step relaxation pattern familiar from other correlation functions
close to the glass transition.  The non-linear shear thinning is clearly
seen as a speeding up of the final relaxation time in $G(t)$ from the
equilibrium case ($\tau\approx10^3$) to that of steady-shear, for which
the characteristic time scale is $\dot\gamma^{-1}$.  A plateau at times
around $t\approx1$ is still seen as a remnant of the cage effect. Note
that only much closer to the glass transition would a well-established
constant plateau develop, defining the solid shear modulus $G_{\infty}$.
For the parameters chosen here, the modulus $G(t)$ still shows a finite slope
in the plateau regime. Considering that
$\sigma(t)$ is the time-integral over $G(t)$ leads to the conclusion
that the `elastic' small-strain regime visible in Figs.~\ref{fig2} and
\ref{joe1b} is not truly linear in $\gamma$.  Indeed, the simulation
data could be fitted with a power law $\sigma\propto\gamma^x$ with
some effective exponent $x<1$ in this time window.

\begin{figure}
\begin{center}
\includegraphics[width=.55\textwidth]{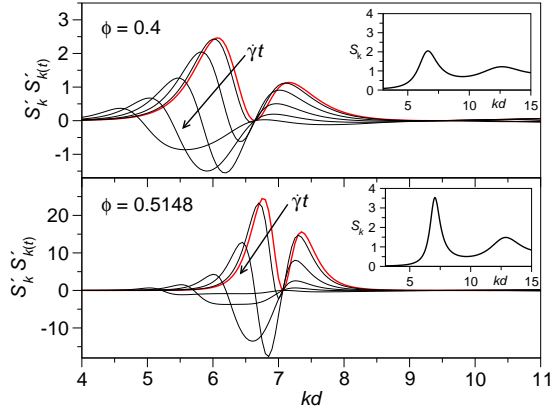}
\caption{
The product $S'_k S'_{k(t)}$ which enters the stress vertex and which leads to 
negative values of the modulus $G(t)$ at long times. 
The uppermost panel and inset show $S'_k S'_{k(t)}$ and $S_k$, respectively, 
for a packing fraction $\phi=0.4$. 
Curves are for strain values $\dot\gamma t=0$, $0.1$, $0.2$, $0.3$, $0.4$
and $0.5$ (where the arrow indicates the direction of increase).
The lower panel shows the same quantities for a packing fraction of $\phi=0.5148$, 
corresponding to the packing fraction used in our comparison with simulation. 
The five values of the strain considered are marked in Fig.~\ref{joe2b}.
\label{structure}}
\end{center}
\end{figure}
The transient shear modulus in addition shows a characteristic
negative dip for strains around $\dot\gamma t\approx0.1$, which
for the highest shear rate corresponds to $tD_0/d^2\approx 20$.
From the integral for the stress, Eq.~\eqref{stress}, it is
apparent that an overshoot in $\sigma(t)$ is the result of such a negative
portion in $G(t)$ for long times. Recall that $G(t)$ is a stress-stress
autocorrelation function, so that this negative dip may be interpreted
as a `backscattering' of stresses.  It arises from the final stage of
the relaxation, commonly referred to as the $\alpha$ process, which is
attributed to the breaking up of nearest-neighbour cages.  As a result
MCT does not correlate the stress overshoot to peculiar inhomogeneities,
such as localized shear events.

Within our MCT-based approach negative values of the modulus occur as a result 
of the factor $S'_k S'_{k(t)}$ appearing in Eq.(\ref{modulus}). 
The derivative $S'_k$ oscillates about zero as a function of $k$ and attains 
maximal positive and negative values in the vicinity of the first peak in $S_k$. 
As the strain is increased the resulting de-phasing of the two factors $S'_k$ and 
$S'_{k(t)}$ leads to negative regions in the product $S'_k S'_{k(t)}$. 
This behaviour is shown in more detail in Fig.\ref{structure} for two values 
of the packing fraction $\phi$, where the structure factors are taken from the 
analytic Percus-Yevick theory.
Comparing the curves for the two densities it is apparent that 
the first peak in $S_k$ is considerably narrower and steeper for $\phi=0.5148$ 
than for $\phi=0.4$, leading to larger values of $S'_k$ in this region.
In each case the product $S'_k S'_{k(t)}$ becomes increasingly negative 
as the strain is increased from $\dot\gamma t=0.1$ to $\dot\gamma t=0.2$. At longer times, the decay of the correlators in Eq.(\ref{modulus}) suppresses the negative vertex.
The inset to Fig.~\ref{joe2b} shows the modulus $G(t)$ as a function of strain. 
For strains in the range $0.1$ to $0.25$ the modulus attains its most negative 
values.
Comparison of the two panels in Fig.~\ref{structure} shows that for $\phi=0.5148$ the 
negative region in $S'_k S'_{k(t)}$ is much more developed at 
$\dot\gamma t=0.2$, 
relative to $\phi=0.4$. This reflects the differences in the first peak of 
the structure factor at the two densities and suggests that systems with 
a steeper and narrower first peak in $S_k$ will possess a more pronounced 
overshoot within the present MCT-based approach. 
Structure factors with a steeper and narrower first peak are known to 
occur in two-dimensional systems \cite{bayer07}.

\begin{figure}
\begin{center}
\includegraphics[width=.55\textwidth]{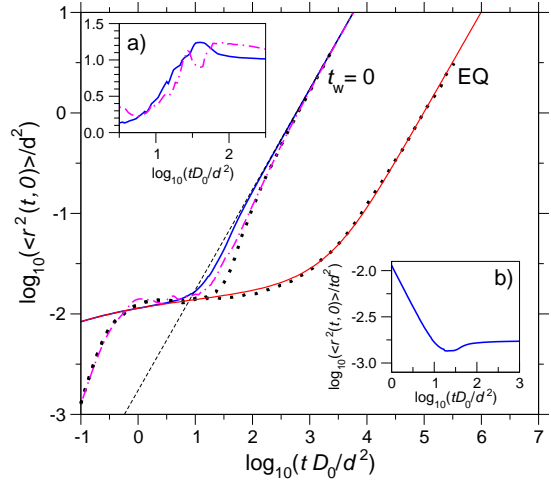}
\caption{
Comparison of the mean-square displacements from simulation (vorticity
direction, dots) and theory (solid lines), for equilibrium (EQ) and for
the highest shear rate considered in the simulations (waiting time
$t_{\rm w}=0$). A superdiffusive regime is apparent in the theoretical MSD
for $t_{\rm w}=0$, establishing the connection to the stress overshoot in
Fig.~10. Inset (a) shows the effective exponent from the theory for
$t_{\rm w}=0$ obtained from the logarithmic derivative. The peak is at a
value of $\sim 1.2$. Inset (b) shows the MSD divided by $t$. Dash-dotted
lines correspond to the MSD calculated from the simulated $\sigma(t)$
at $\dot\gamma=3\times10^{-3}$ and the GSE relation (see text). The dashed
line indicates the linear diffusive behavior in steady state.
\label{joe3b}}
\end{center}
\end{figure}
We have now established two features of the transient dynamics immediately
following the instantaneous switch-on of shear, both of which arise
at strains of about $10$\% of the particle diameter.  The first of
these, $\sigma(t)$ (equivalently $G(t)$), concerns the collective
dynamics, while the other, $\langle r^2(t,t_{\rm w})\rangle$, concerns the
single-particle dynamics of a tracer particle.  It is intriguing to see
in which way these conceptually different pieces of information might
be related and to connect the macroscopic rheological information
contained in $\sigma(t)$ with the microscopic particle motion. We
consider in more detail the transient MSD, $\delta r^2(t) = \langle
r^2(t,t_{\rm w}=0)\rangle$.  An idea that has proven particularly successful
in the quiescent system is that of a generalized Stokes-Einstein (GSE)
relation, proposed by Mason and Weitz as a correspondence between macro-
and micro-rheology \cite{mason}. Within this approach the diffusion
coefficient $D$ is related to the viscosity $\eta$ by
\begin{equation}
  D\approx\frac{k_BT}{3\pi d\alpha\eta}=
      \frac{k_BT}{3\pi d\alpha\int_0^\infty ds\,G(s)}\,.
\label{gse}
\end{equation}
In the context of quiescent MCT, this relation exemplifies the coupling
of the $\alpha$-relaxation times \cite{goetze92}, and even quantitatively
works surprisingly well for the quiescent hard-sphere system
\cite{fuchsmayr}. Quantitatively accurate results can be obtained if
one allows for a small correction to the geometrical prefactor $3\pi d$
appearing in the original Stokes-Einstein relation, $\alpha\neq1$.

In addition to the correlator $\Phi_k(t)$, MCT also yields equations
of motion for the tagged-particle correlation functions, which yield
an equation for the MSD in the $q\to0$ limit.  Taking the form of
the equation of motion for $\delta r^2(t)$ from the quiescent theory
(ignoring any generalizations that come about due to the anisotropic
flow geometry), we can write
\begin{equation}
\delta r^2(t) + \frac{D_0d}{k_BT}\!\int_0^t \!\!dt'\, 
m^{(s)}(t-t')\delta r^2(t') = 6D_0\,t,
\label{msd}
\end{equation}
where $m^{(s)}(t)$ is the tagged-particle memory kernel evaluated as
$q\to0$. An analysis of the above equation at large times immediately
gives an expression for the diffusion coefficient in terms of the
memory kernel,
\begin{equation}
D=\frac{D_0}{1+D_0\int_0^\infty ds\,m^{(s)}(s)}\approx
\frac1{\int_0^\infty ds\,m^{(s)}(s)}\,.
\label{longtime}
\end{equation}
Comparison of (\ref{longtime}) with (\ref{gse}) clearly establishes
a relationship between the time integrals of $ m^{(s)}(t)$ and
$G(t)$. However, the connection is deeper than simple equality of
the integrated quantities. Studies performed using the quiescent
mode coupling theory have established that the MCT approximation to
$G(t)$ is in excellent agreement (up to a constant prefactor) with
the MCT approximation to $m^{(s)}(t)$ for the hard sphere system
\cite{fuchsmayr}.  In this work we make the assumption that this
correspondence holds also in the non-linear regime for which $G(t)\equiv
G(t,\dot\gamma)$.  Given this assumption we obtain a direct correspondence
between the MSD and the derivative of the shear stress after startup,
$(d/dt)\,\sigma(t)\propto G(t)$,
\begin{equation}
m^{(s)}(t)\approx \left(\frac{d}{k_BT}\right) 3\pi\,\alpha\, G(t)
  =\left(\frac{d}{k_BT}\right)\frac{3\pi\,\alpha}{\dot\gamma}\,
    \frac{d}{dt}\sigma(t)\,.
\label{mw}
\end{equation}

We now consider application of the GSE approximation (\ref{mw}) to
the calculation of the mean-squared displacement using (\ref{msd}).
In Fig.~\ref{joe3b} we compare the $t_{\rm w}=0$ curves obtained
from simulation and theory for the largest shear rate considered in
the simulation.  Also shown is the equilibrium curve that has been used
to adjust $\Delta\phi$ in fitting the equilibrium simulation result. At
short times differences arise due to the different microscopic dynamics,
as expected, but for the long-times relevant to our discussion, the
fit is convincing.  The MCT-GSE theory displays a super-diffusive regime
at intermediate times, qualitatively reproducing the phenomenology of
the simulation.  In order to quantify this superdiffusivity, we show in
inset (a) the logarithmic derivative of the MSD, yielding an effective
exponent for $\delta r^2(t)$.  While in the simulation this quantity
grows up to $2$ (reflecting ballistic motion) in the intermediate regime,
our MCT calculation produces effective exponents up to roughly $1.2$.
This situation is much closer to the results of our colloidal experiments.

Note that from (\ref{longtime}), a super-diffusive regime in the MSD
can only arise if $m^{(s)}(t)$ takes on negative values.  By virtue
of (\ref{mw}), the presence of a super-diffusive regime is therefore
intimately connected to the stress overshoot and the corresponding
overrelaxation feature found in $G(t)$.  It is therefore tempting to
assume that the more pronounced super-diffusive regime found in the
simulations relative to the theory is a consequence of the correspondingly
stronger stress overshoot (see Fig.~\ref{joe1b}).  In order to investigate
this connection further we have directly calculated the mean-squared
displacement from the simulation curve for $\sigma(t)$, using (\ref{mw})
to obtain the memory function $m^{(s)}(t)$.  In this calculation,
the Newtonian-dynamics equivalent of Eq.~\eqref{msd} was used with a
prefactor $\alpha=0.833$ (accounting for the numerical uncertainty that
arises from the neccessary smoothening of the derivative of the simulated
$\sigma(t)$).  The resulting MSD is shown in Fig.~\ref{joe3b} (dash-dotted
lines).  For times longer than $tD_0/d^2\approx 100$ the MSD obtained in
this fashion closely follows the directly simulated quantity.  However,
for shorter times the agreement is less satisfactory and significant
deviations are visible.  The logarithmic derivative shown in inset of
Fig.~\ref{joe3b}a confirms this. As a result of these findings we conclude
that the approximation (\ref{mw}) remains qualitatively reliable in the
non linear regime but that it fails quantitatively in the super-diffusive
time window where the stress overshoot is strongly pronounced.

\begin{figure}
\begin{center}
\includegraphics[width=.55\textwidth]{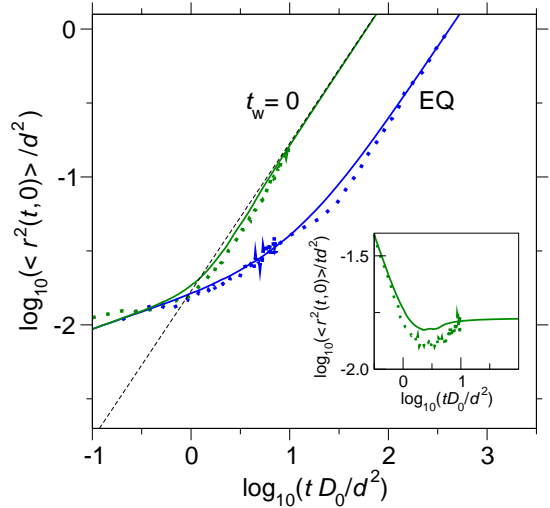}
\caption{Theoretical fits to the experimental mean-square displacement 
(dotted) for equilibrium and $t_{\rm w}=0$ (solid lines) using $\alpha=0.95$
(see text). The inset shows the MSD divided by $t$ and highlights the
super-diffusive regime. \label{mct_exp}} 
\end{center}
\end{figure} 
Finally, we consider application of the GSE-MCT relation to fitting
the experimental data for the MSD, the results of which can be seen in
Fig.~\ref{mct_exp}.  We follow the same procedure as when fitting the
simulation data and consider first the equilibrium MSD. By adjusting
$\Delta\phi$ we fix the ratio of final to microscopic relaxation time and
find a value $\Delta\phi=-1.1\times 10^{-2}$ reproduces the experimental
results to a satisfactory level, using also a value $\alpha=0.95$ little different from unity.  We next increase the shear rate until
the long time diffusion matches that of the simulation. This is achieved
for ${\rm \We}=16$ (corresponding to $\dot\gamma=5\times 10^{-2} D_0/d^2$).  We find
a super-diffusive regime in qualitative agreement with the experiment at ${\rm \We}=7$, but
which is somewhat underestimated in magnitude. This lends support to our argument
that the approximation (\ref{mw}) is less reliable in the super-diffusive
regime than in equilibrium. In the inset we show the MSD divided by $t$
which confirms the relatively weaker superdiffusion.

\section{Conclusions}\label{sec:conclusions}
We have studied the transient dynamics of a glassforming system that
is subjected to the sudden commence of steady shear flow. Results from
computer simulation, confocal-microscopy experiment, and mode-coupling
theory give a consistent picture that reveals peculiar dynamical features
around the time where the system reaches $10\%$ strain for the first
time. In particular, the mean-squared displacements in both the vorticity
and the gradient direction show super-diffusive behaviour at intermediate
times, interpolating between the equilibrium curve that is followed for
$\gamma\lesssim0.1$ and the faster steady-state curve that is followed
for $\gamma\gtrsim1$. The super-diffusive behaviour is more pronounced
in the (ballistic) simulation, giving rise to effective exponents of $2$,
and less pronounced in the colloidal experiment.

Again at strains of $\gamma\approx0.1$, a well-recognized feature in
the collective dynamics is found, visible in the building up of the
shear stress as a `stress overshoot' (a maximum in $\sigma(t)$ followed
by a decrease towards a smaller steady-state value at later times).
While we cannot propose a simple picture of microscopic motion behind
this stress overshoot, we are able within MCT to directly relate it to
a `stress overrelaxation', visible in the corresponding stress-stress
autocorrelation function in the late $\alpha$-relaxation regime. It
can thus be attributed to the peculiar way in which cages break up, and
MCT predicts its specific shape and magnitude to be system-dependent. We
have demonstrated that MCT in this way qualitatively describes the stress
overshoot seen in the simulation, albeit quantitatively underestimated.

A generalized Stokes-Einstein relation can be used to relate the above
two features of the transient collective and single-particle dynamics. We
have used MCT in combination with this close link between micro- and
macro-rheology in order to demonstrate that the theory-predicted stress
overrelaxation quite naturally leads to a super-diffusive regime in
the mean-squared displacement. Again, its magnitude is less pronounced
than in the simulation, but compares reasonably well with experiment,
so that the remaining differences can be attributed to the microscopic
difference in the systems, the isotropic nature of our approximations,
and violations of the generalized Stokes-Einstein relation.

\ack{
We thank K. Binder for useful discussions. 
We thank R. Besseling for providing routines for calculation of the
velocity profile, A. Schofield for providing PMMA particles and D. Vobis
for help with the design and construction of the shear cell.
We thank SFB TR6 for support. 
Th.~V.\ thanks for funding through the Helmholtz-Forschungsgemeinschaft
(Impuls- und Vernetzungsfonds, VH-NG-406). We acknowledge a substantial 
grant of computer time at the J\"ulich multiprocessor system
(JUMP) of the John von Neumann Institute for Computing (NIC).}

\bibliography{lit2}
\bibliographystyle{iopart-num}
%
%
%
%
%
%
%

\end{document}